\documentclass[reprint,superscriptaddress,nofootinbib,amsmath,amssymb,aps]{revtex4-1}

\usepackage{graphicx}
\usepackage{dcolumn}
\usepackage{bm}
\usepackage[colorlinks=true, allcolors=blue]{hyperref}
\usepackage{color}

\def\mnras{Mon. Not. R. Astron. Soc. }

\def\apj{Astrophys. J.}
\def\apjl{Astrophys. J. Lett.}
\def\prd{Phys. Rev. D}
\def\prl{Phys. Rev. Lett.}

\def\aap{Astron. Astrophys.}

\def\nat{Nature}

\begin{document}

\preprint{APS/123-QED}

\title{Binaries Wandering Around Supermassive Black Holes Due To
Gravito-electromagnetism}

\author{Xian Chen}
\email{Corresponding author: xian.chen@pku.edu.cn}
\affiliation{Department of Astronomy, School of Physics, Peking University, 100871 Beijing, China}
\affiliation{Kavli Institute for Astronomy and Astrophysics at Peking University, 100871 Beijing, China}

\author{Zhongfu Zhang}
\affiliation{Astronomy Department, School of Physics, Peking University, 100871 Beijing, China}

\date{\today}

\begin{abstract}
Extreme-mass-ratio inspirals (EMRIs) are important sources for space-borne
gravitational-wave (GW) detectors.  Such a source normally consists of a
stellar-mass black hole (BH) and a Kerr supermassive BH (SMBH), but recent
astrophysical models predict that the small body could also be a stellar-mass
binary BH (BBH).  A BBH reaching several gravitational radii of a SMBH will
induce rich observable signatures in the waveform, but the current numerical
tools are insufficient to simulate such a triple system while capturing the
essential relativistic effects.  Here we solve the problem by studying the
dynamics in a frame freely falling alongside the BBH.  Since the BBH is
normally non-relativistic and much smaller than the curvature radius of the
Kerr background, the evolution in the free-fall frame reduces to essentially
Newtonian dynamics, except for a perturbative gravito-electromagnetic (GEM)
force induced by the curved background.  We use this method to study the BBHs
on near-circular orbits around a SMBH and track their evolution down to a
distance of $2-3$ gravitational radii from the SMBH.  Our simulations reveal a
series of dynamical effects which are not shown in the previous studies using
conventional methods. The most notable one is a radial oscillation and azimuthal
drift of the BBH relative to the SMBH.  These results provide new insight into
the evolution and detection of the EMRIs containing BBHs.  
\end{abstract}

\maketitle

\section{Introduction}\label{sec:intro}

The idea that two stellar-mass compact objects ($\sim10M_\odot$)  could form a
binary and merge near a supermassive black hole (SMBH, $\gtrsim10^6M_\odot$) is
an old one. It could date back to the 1970s, soon after Weber reported a strong
signal in his gravitational-wave (GW) experiment \cite{weber70}. Several
authors proposed that the signal could come from a GW source, such as a merging
binary, near the innermost stable circular orbit (ISCO) around the SMBH in the
Galactic Center \cite{campbell73,lawrence73,ohanian73}. The signal appeared
strong because it was gravitationally lensed by the SMBH.  This idea, however,
went largely unnoticed since the controversy of Weber's result.

The recent observations of GWs by the Laser Interferometer Gravitational-wave
Observatory (LIGO) and the Virgo detectors has revived the interest in the very
same idea.  So far, LIGO and Virgo have detected nearly $100$ binary black
holes (BBHs) \cite{GWTC3}.  The majority of them seem significantly more
massive than the black holes (BHs) found previously in X-ray binaries.  The
sharp contrast raises an interesting question about the origin of these massive
objects.  More than one astrophysical channels may be responsible for producing
them (see \cite{2016ApJ...818L..22A,2019ApJ...882L..24A,2020ApJ...900L..13A}
for summaries). Among the many possibilities is a BBH orbiting around and
dynamically interacting with a SMBH.

On one hand, the SMBH could accelerate the merger of the BBH by exciting its
eccentricity, due to either the Von Zeipel-Lidov-Kozai mechanism when the
mutual inclination of the triple system is high
\cite{antonini_perets_2012,petrovich_antonini_2017,hoang18}, or the evection
resonance when the inclination is small \cite{liu20,bhaskar22,munoz22}.
In either case,
the post-merger BH is likely retained in the vicinity of the SMBH due to the
large escape velocity, and hence could participate in another
merger \cite{MillerLauburg09}. 
The repeated  mergers could explain the large BH mass detected by
LIGO/Virgo.  On the other, if the SMBH has an accretion disk, as would be the
case in an active galactic nucleus (AGN), 
a BH embedded in the disk could grow by accreting the surrounding gas.
More importantly, BBHs could form and merge more
frequently in the assistance of the gas 
\cite{baruteau11,mckernan_ford_2012,bartos_kocsis_2017,stone_metzger_2017}. 
Interaction  
with the other compact objects in the accretion disk may further enhance
the merger rate of BBHs
\cite{2018MNRAS.474.5672L,2019ApJ...876..122Y}.  
It has been shown that
the merger rate
could be comparable to the LIGO/Virgo event rate
in the above astrophysical scenarios
\cite{fragione19,tagawa_haiman_2019,ArcaSedda20,grobner20,ford21mckernan,gayathri21,zhang21,samsing22}.

The GW signal from such a BBH orbiting a SMBH is the focus of many recent
studies.  It has been shown that the GWs emitted from the BBH will be
periodically modulated by the Doppler shift
\cite{inayoshi_tamanini_2017,meiron_kocsis_2017,wong_baibhav_2019,tamanini20}
as well as an aberrational effect \cite{Torres-OrjuelaEtAl2020}. This
modulation may be negligible for a LIGO/Virgo event, because the signal is too
short ($<1$ s) compared with the orbital period around the SMBH, therefore does
not allow the velocity of the center-of-mass (CoM) of the BBH to substantially
change \cite{ChenAmaro-Seoane2017}. However, the effect is detectable by the
space-borne Laser Interferometer Space Antenna (LISA), because LISA could track
a BBH for as long as several years \cite{Sesana2016}.  The long waveform can
also reveal the variation of the internal eccentricity of the BBH, which is
induced by the aforementioned Von Zeipel-Lidov-Kozai cycle
\cite{meiron_kocsis_2017,hoang19,randall19,gupta20,chandramouli22}.  If the
SMBH is spinning, the frame-dragging effect will further perturb the BBH and
leave a detectable imprint in the GW signal 
\cite{fang19huang,liu19spin}.

These earlier studies normally assume that the BBH is hundreds to thousands of
gravitational radii away from the SMBH.  This assumption is partly motivated by the
earlier theoretical predictions (e.g.
\cite{antonini_perets_2012,mckernan_ford_2012}) and partly due to a practical
reason, that is the relativistic effect at such intermediate distances is
mild so that it can be modeled with the post-Newtonian (PN) formalism
\cite{will14}.  However, more recent studies suggest that the BBH could reach a
distance as small as ten gravitational radii from the SMBH
\cite{2018CmPhy...1...53C,addison_gracia-linares_2019,peng21}. Such a system
resembles in many ways an important category of LISA sources, known as 
extreme-mass-ratio inspirals (EMRIs) \cite{Amaro-SeoaneEtAl07}.  But the
distinction is also clear. Instead of having only one stellar-mass BH, now the
system contains two stellar-mass BHs,  bound by their self-gravity.

As the distance between the BBH and the SMBH decreases,
interesting phenomena emerge, which have attracted intense
scrutiny lately.
(i) The Doppler and
gravitational redshift becomes so strong that in the detector frame the BBH
will appear more massive and more distant \cite{2019MNRAS.485L.141C}. The same
effect can also lower the observed GW frequency of a binary neutron star and
make its detection easier for ground-based observatories \cite{vijaykumar22}.
(ii) The large spacetime curvature around the SMBH could
bend the null geodesics of the GWs emitted by the BBH, producing  
lensed images \cite{dorazio20,yu21}, echoes \cite{gondan21}, as well as a
Shapiro delay \cite{sberna22}. If the
SMBH is spinning and the BBH lies in the equatorial plane, the GWs leaving the
BBH may be amplified by a Penrose-like process \cite{gong21}.  (iii) When
the frequency of the GWs from the BBH matches a quasi-normal mode of
the SMBH, the SMBH could be resonantly excited \cite{cardoso21}.  (iv) The fast
motion of the BBH around the SMBH can distort the pattern of its GW radiation 
due to aberration, which will induce additional
higher modes in the GW signal
\cite{torres-orjuela_chen_2020b}.  (v) Since the coalescence of the BBH
produces high-frequency GWs ($10-10^2$ Hz) and its motion around the SMBH
generates low-frequency ones ($\sim10^{-3}$ Hz), the triple system at its final
evolutionary stage becomes a multi-band GW source. Such a source is ideal for
measuring the energy and momentum carried away by GWs, as well as
constraining the mass of gravitons \cite{han_chen_2018}.

Despite the increasing interest in considering a smaller separation between the
BBH and the SMBH, the numerical tools we have today may not be adequate to
model the dynamics of such a triple system.  The commonly used PN approximation
breaks down as the separation shrinks to the order of ten gravitational radii
because the velocity of the BBH relative to the SMBH becomes a significant
fraction of the speed of light.  A more subtle issue is that many earlier
studies, in order to simplify the computation of the waveform, treated the BBH
as a single body and modeled its orbit around the SMBH with a geodesic line
(e.g., \cite{gupta20,chandramouli22}). The validity of this treatment has not
been tested when the separation is so small. In the end, a binary has internal
structures and for this reason differs from a test particle. 

One possible solution is to take advantage of the equivalence principle and
investigate the problem in a frame freely falling together with the BBH.
Especially when the BBH is non-relativistic, the dynamics in this free-fall
frame (FFF) is essentially Keplerian, except for a perturbing force induced by
the curvature of the SMBH background. We notice that a similar method has been
used to model a binary star close to a Schwarzschild SMBH \cite{komarov18},
though in their work the authors implicitly assumed that the CoM of the binary
follows a geodesic line.  Here we revise their method to simulate the evolution
of a BBH down to a distance of $2-3$ gravitational radii from a Kerr SMBH and,
at the same time, self-consistently track  the CoM of the BBH which turns out
to be non-geodesic.

The paper is organized as follows.  In Section~\ref{sec:label} we describe the
theoretical framework of simulating the evolution of the BBH in its FFF. Based
on the observation that the perturbation by the Kerr background induces
gravito-electromagnetic (GEM) forces in the FFF, we argue that these GEM forces
will drive the CoM of the BBH away from a geodesic line.  In
Section~\ref{sec:num} we carry out numerical simulations to test our
prediction.  We also analyse the evolution in different frames to better
understand the cause and the effects of the geodesic deviation.  In
Section~\ref{sec:initial} we vary the initial conditions to showcase the
richness of the dynamical effects induced by the close interaction between the
BBH and the SMBH.  Based on our simulation results, we discuss in
Section~\ref{sec:obs} the possible observational signatures imprinted in the GW
signal of such a triple system.  Finally, we conclude in Section~\ref{sec:con}
and point out several caveats of the current work, as well as their possible
solutions.  Throughout the paper, we use geometrized units where $G=c=1$. 

\section{Theory}\label{sec:label}

\subsection{Equation of motion}

The system of our interest consists of a SMBH with a mass of $M$ ($10^6\lesssim
M\lesssim10^9$) and a BBH within a distance of $r\lesssim10M$ from the SMBH.
Following the convention, we refer to the self-gravitating BBH as the ``inner
binary'' and the trajectory of its CoM around the SMBH as the ``outer orbit''.
We consider a rotation of the SMBH and denote the dimensionless spin parameter
as $s$.  We further denote the masses of the two BHs of the BBH as $m_1$ and
$m_2$. Then the total mass of the inner binary is $m_{12}=m_1+m_2$, which is
typically tens of  solar masses. 

Although the outer orbit is highly relativistic, the inner one is much simpler.
On one hand, the semimajor axis of the inner binary, which we denote by $a$, is
$10^2-10^3$ times smaller than $r$
\cite{2018CmPhy...1...53C,addison_gracia-linares_2019,peng21}, and hence also
much smaller than the curvature radius of the background Kerr metric. This fact
indicates that the spacetime is sufficiently flat sufficiently close to the
BBH.  On the other hand, $a$ is typically $10^3-10^4$ times greater than
$m_{12}$, so that the relative speed between the two small BHs
($\sim\sqrt{m_{12}/a}$) is non-relativistic. 

Given the above hierarchy, it is the most convenient to use the Fermi normal
coordinates to describe the inner orbit \cite{komarov18}.  They are the
coordinates a free-fall observer would naturally choose since sufficiently
close to the observer the metric is approximately Minkowskian.  We denote these
coordinates as $(\tau, \mathbf{x})=(\tau, x, y, z)$, where the origin of the
spatial coordinates coincides with the location of the observer\footnote{Here
we use $\tau$ to label the time in the free-fall frame, and we save the symbol
$t$ to denote the time in the Boyer-Lindquist coordinate.}.  In such a frame,
the perturbation to the Minkowski metric by the SMBH is of the order of
$(\mathbf{x}/r)^2$, which is small in our case of $|\mathbf{x}|\simeq a\ll r$.
We choose the origin of the FFF (the free-fall observer) to coincide with and
has the same velocity as the CoM of the BBH.  Then the initial velocities of
the BHs in the FFF are Keplerian, and our problem reduces essentially to
Newtonian dynamics.  We note that the origin of the FFF by construction follows
a geodesic line, but the CoM of the BBH does not necessarily follow the same
geodesic, as we will show later.

The aforementioned quadratic-order perturbation to the Minkowskian metric 
induces GEM forces in the FFF \cite{mashhoon03}.
These forces can be formulated with
\begin{equation}
	\mathbf{F}=-m\mathbf{E}-2m\mathbf{v}\times \mathbf{B}, \label{eq:F}
\end{equation}
where 
$m$ is the rest mass of an object,
$\mathbf{v}:=d\mathbf{x}/d\tau$ is its
velocity in the FFF, and
$\mathbf{E}$ and $\mathbf{B}$ are, respectively,
the gravito-electric and gravito-magnetic fields.
The minus signs
in the above equation reflect the fact that mass has a negative charge in gravito-electromagnetism.
To first order in $\mathbf{x}$, the components of 
the GEM fields can be calculated with
\begin{eqnarray}
	E_i(\tau,\mathbf{x})&=&R_{0i0j}(\tau)x^j,\label{eq:E}\\
	B_i(\tau,\mathbf{x})&=&-\frac{1}{2}\epsilon_{ijk}{R^{jk}}_{0l}(\tau)x^l,\label{eq:B}
\end{eqnarray}
where $R_{0i0j}$ and ${R^{jk}}_{0l}$ are the components of the Riemann tensor, $\epsilon_{ijk}$ is the
Levi-Civita symbol, and $i$, $j$, $k$, and $l$ are spatial indices which take
the values $1, 2, 3$.

Combining the Newtonian and GEM forces, we can 
write the equation of motion in the FFF as
\begin{equation}
	m_a\frac{d^2\mathbf{x}_a}{d\tau^2}=-m_a m_b\frac{\mathbf{x}_a-\mathbf{x}_b}{|\mathbf{x}_a-\mathbf{x}_b|^3}+\mathbf{F}_a(\tau,\mathbf{x}_a,\mathbf{v}_a),
\label{eq:eom}
\end{equation}
where $a, b=1,2$, denoting the two stellar-mass BHs.  We note that the last
equation is valid even when the CoM of the BBH deviates from the origin of the
FFF. The validity of this equation only requires that the position of the CoM,
$\mathbf{x}_{\rm CoM}=(m_1\mathbf{x}_1+m_2\mathbf{x}_2)/m_{12}$, satisfies the
condition $|\mathbf{x}_{\rm CoM}|\ll r$. Otherwise, if the BBH wanders too far
away from the origin of the FFF, the perturbation term $(\mathbf{x}/r)^2$ in
the metric will no longer be small and the formulae for the GEM fields will be
invalid.

\subsection{Calculating the gravito-electromagnetic forces}\label{sec:force}

Although it is the simplest to write the equation of motion in the FFF, 
it
is not straightforward to calculate the GEM fields in the same frame.  This is
caused by the fact that the Riemann tensor is conventionally derived in the
``locally non-rotating frame'' (LNRF \cite{bardeen72}), a frame not in free
fall. Here we will show that the FFF in general differs from the
LNRF by a boost and a rotation. Based on this understanding, we will derive the
GEM forces in the FFF.

We start from the well-known
Boyer-Lindquist coordinates $(t,r,\theta,\phi)$ in which the metric is
\begin{eqnarray}
	ds^2&=&-(1-2Mr/\Sigma)dt^2-(4Mar\sin^2\theta/\Sigma)dtd\phi+(\Sigma/\Delta)dr^2\nonumber\\
	&+&\Sigma d\theta^2+(r^2+a^2+2Ma^2r\sin^2\theta/\Sigma)\sin^2\theta d\phi^2,
\end{eqnarray}
and the functions $\Delta$, $\Sigma$, and $A$ are defined by
\begin{eqnarray}
\Delta&:=&r^2-2Mr+a^2,\nonumber\\
\Sigma&:=&r^2+a^2\cos^2\theta,\nonumber\\
A&:=&(r^2+a^2)^2-a^2\Delta\sin^2\theta.
\end{eqnarray}
The bases of the Boyer-Lindquist coordinates are not orthonormal.  Therefore,
it is not easy to derive the Riemann tensor in this frame.

On the
contrary, the LNRF has orthonormal coordinates. It is the frame used by an
observer whose world line
follows constant $r$ and $\theta$ but different
$\phi$, when viewed from the Boyer-Lindquist coordinates.  The coordinates of the LNRF, $(X^t,
X^r, X^\theta, X^\phi)$, is related to the Boyer-Lindquist coordinates as
\begin{eqnarray}
	dX^t&=&\left(\Sigma\Delta/A\right)^{1/2}dt,\nonumber\\ 
	dX^r&=&\left(\Sigma/\Delta\right)^{1/2}dr, \nonumber\\ 
	dX^\theta&=&\Sigma^{1/2}d\theta, \nonumber\\
	dX^\phi&=&-\frac{2Mar\sin\theta}{(\Sigma A)^{1/2}}dt+\left(\frac{A}{\Sigma}\right)^{1/2}\sin\theta d\phi.
\end{eqnarray}
It has been shown that the Riemann tensor is much simpler to derive in the LNRF
than in the Boyer-Lindquist coordinates \cite{bardeen72}.  In the following, we
denote the Riemann tensor in the LNRF as $R_{\mu\nu\rho\sigma}$.

To transform $R_{\mu\nu\rho\sigma}$ into the FFF, we notice that at any
instance in time, the local frame of a free-fall observer differs from the LNRF
at the observer's location by a Lorentz transformation \cite{bardeen72}.
Therefore, given the velocity of the FFF relative to the LNRF, we can in
principle derive the Riemann tensor in the FFF by a Lorentz transformation.
However, because the velocity in general is not aligned with any of the spatial
axes of the LNRF, the Lorentz transformation will be a complicated one.

In practice, we can simplify the Lorentz transformation according to the
problem we are dealing with.  We notice that BBHs could be delivered to the
vicinities of spinning SMBHs along near-circular orbits in the equatorial
planes of the SMBHs. This situation has been predicted by the previous models related to
AGNs \cite{peng21}. In this case, it is natural to choose a free-fall
observer on a circular orbit in the equatorial plane.
Viewed from the LNRF, this observer is moving in the azimuthal ($X^\phi$) direction. The speed is
\begin{equation}
	u=\frac{\pm M^{1/2}(r\mp2aM^{1/2}r^{1/2}+a^2)}{\Delta^{1/2}(r^{3/2}\pm aM^{1/2})},
\end{equation}
where the upper signs refer to prograde orbits and the lower ones refer to
retrograde orbits \cite{bardeen72}.   
Given these conditions, the simplest coordinate system the free-fall observer can use
to calculate the Riemann tensor is such that  
one spatial axis is in the same direction as the velocity
($X^{\phi}$ direction), and the other two are aligned, respectively, with the $X^r$ and
$X^\theta$ axes of the LNRF. Then the Lorentz transformation matrix
has the simple form
\begin{equation}
	\Lambda_{\mu'}^{\mu} =
	\left( {\begin{array}{cccc}
	  \gamma & 0 & 0 & \gamma\beta \\
	  0      & 1 & 0 & 0 \\
	  0      & 0 & 1 & 0 \\
	  \gamma\beta & 0 & 0 & \gamma \\
	\end{array} } \right),\label{eq:LT}
\end{equation}
where $\beta=u/c$, $\gamma=\sqrt{1-\beta^2}$ is the Lorentz factor, and the
primed index $\mu'$ refers to a coordinate in the boosted (observer's) frame.
The transformation of the Riemann tensor,
$R_{\mu'\nu'\rho'\sigma'}=\Lambda_{\mu'}^{\mu}\Lambda_{\nu'}^{\nu}\Lambda_{\rho'}^{\rho}\Lambda_{\sigma'}^{\sigma}R_{\mu\nu\rho\sigma}$,
is also significantly simplified.

However, to appropriately use the Riemann tensor $R_{\mu'\nu'\rho'\sigma'}$, we
must notice a subtle but important difference between the frame where
$R_{\mu'\nu'\rho'\sigma'}$ is derived and the FFF where the equation of motion,
i.e., Equation~(\ref{eq:eom}), applies.  The FFF uses  Fermi normal coordinates
which, by construction, form an inertial frame along the entire geodesic of the
free-fall observer.  However, $R_{\mu'\nu'\rho'\sigma'}$ and the corresponding
GEM fields are calculated in a different frame, which is a Lorentz boost of the
LNRF in a specific direction. Such a boosted frame is an inertia frame only
locally, at one instance in time, but not along the entire geodesic because the
direction of the boost changes with time.  For example, consider a free gyro
placed at the origin of this boosted frame.  As the gyro moves along a geodesic
around the SMBH, its spin axis is fixed in the FFF but in general will precess
relative to the spatial coordinates of the LNRF, as well as the boosted frame.
For this reason, the boosted frame is also known as the ``local inertial
frame'' (LIF).

\begin{figure}[] 
\centering \includegraphics[width=0.45\textwidth]{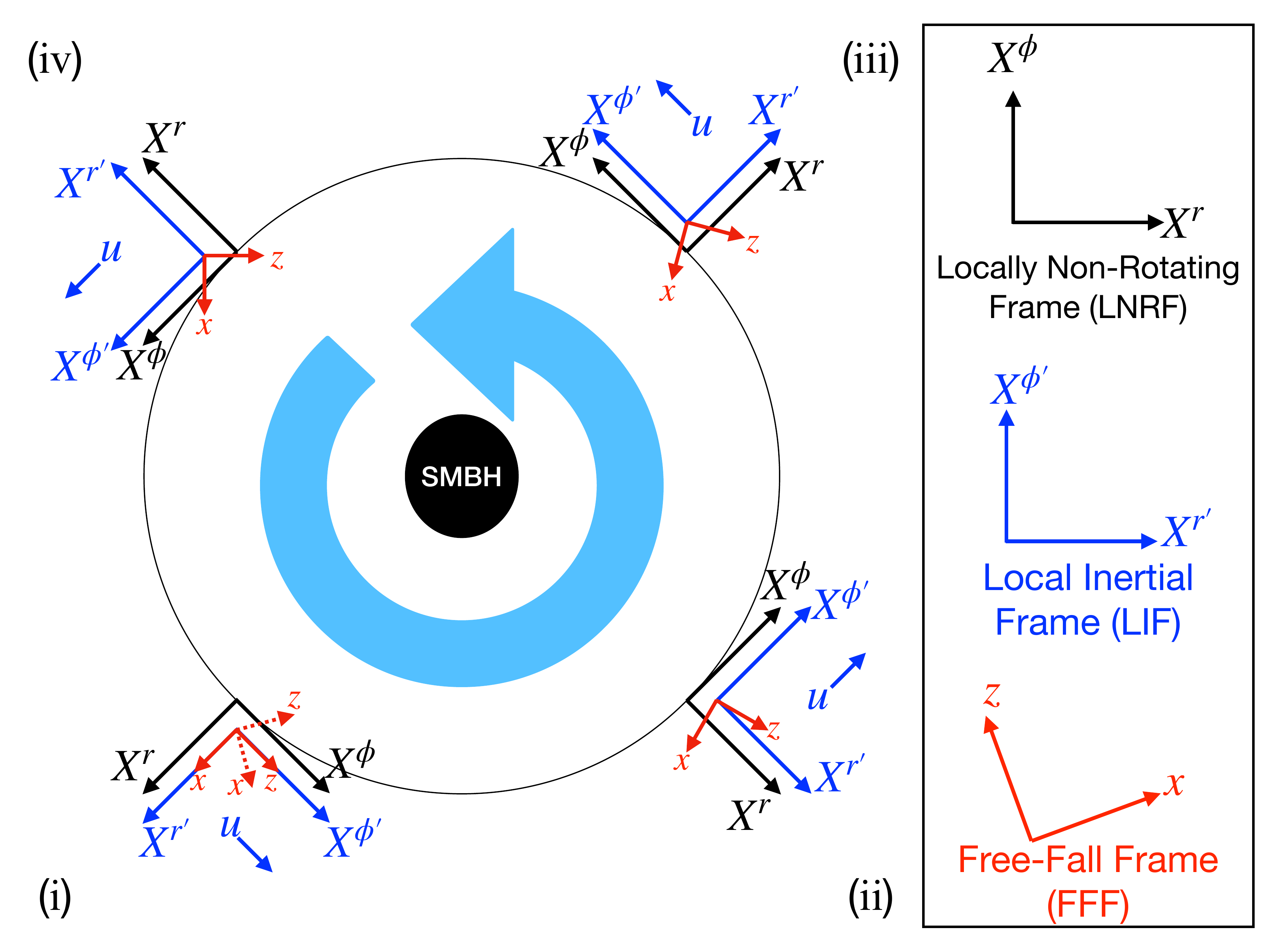}
\caption{Different reference frames used in this paper and their
relative orientation as they rotate around the SMBH. The curved cyan arrow
shows the direction of rotation and the Roman numerals indicate the
sequence of the evolution. The red dotted arrows mark the $x$ and $z$
axes
of the free-fall frame when it has completed one revolution around the
SMBH. The axes misalign with the original ones because of precession.}
\label{fig:frames}
\end{figure}

The relationship between the LNRF, LIF, and FFF is illustrated in
Figure~\ref{fig:frames}, where we have considered a circular orbit for the
free-fall observer. The rotation of the LIF relative to the FFF is clearly
shown. The corresponding angular velocity viewed from the FFF can measured by
the precession of a free gyro, and is $\omega=-\sqrt{M/r^3}$ \cite{rindler90}.
The minus sign indicates that the angular velocity vector ${\boldsymbol\omega}$
is pointing in the $-y$ direction when the free-fall observer is moving on a
prograde orbit around the Kerr SMBH.

Having understood the relationship between the LIF and FFF, we can derive the
GEM forces in the FFF as follows. 
Given the mass $m$, position $\mathbf{x}=(x,y,z)$, and 
velocity $\mathbf{v}$ of an object in the FFF, we first
calculate its position $\mathbf{x'}=(X^{r'},X^{\theta'},X^{\phi'})$ and 
velocity $\mathbf{v}'$ in the LIF with 
\begin{eqnarray}
	X^{r'}&=&x\cos(\omega\tau)-z\sin(\omega\tau),\\
	X^{\theta'}&=&y,\\
	X^{\phi'}&=&x\sin(\omega\tau)+z\cos(\omega\tau),
\end{eqnarray}
and $\mathbf{v}'=\mathbf{v}-{\boldsymbol\omega}\times\mathbf{x}$.
Notice that $\omega<0$ ($\omega>0$) if the FFF is moving along a prograde (retrograde) circular
orbit around the SMBH.
The above transformation reflects the fact that the LIF differs from the FFF by a rotation. 
Using $\mathbf{x}'$ and the Riemann tensor $R_{\mu'\nu'\rho'\sigma'}$,
we can construct the GEM fields in the
LIF, namely, $\mathbf{E}'$ and $\mathbf{B}'$.
Then the GEM force in the LIF, $\mathbf{F}'=(F_{r'},F_{\theta'},F_{\phi'})$, is computed
with $\mathbf{F}'=-m\mathbf{E}'-2m\mathbf{v}'\times\mathbf{B'}$. 
Finally, rotating $\mathbf{F}'$ around the $X^{\theta'}$ axis by
an angle of $\omega\tau$, we get the GEM force in the FFF.
Mathematically, the rotation is performed with 
\begin{eqnarray}
	F_x&=&F_{r'}\cos(\omega\tau)+F_{\phi'}\sin(\omega\tau),\\
	F_y&=&F_{\theta'},\\
	F_z&=&-F_{r'}\sin(\omega\tau)+F_{\phi'}\cos(\omega\tau).
\end{eqnarray}

\subsection{Deviation from free fall}\label{sec:deviation}

We notice that one
assumption which is implicitly adopted by many earlier works is that the outer orbit follows the geodesic of a test
particle (e.g., \cite{fang19huang,liu19spin,gupta20,chandramouli22}).  If this
assumption is valid, we can find a FFF whose origin always coincides with the
CoM of the BBH. We will show 
in this subsection
that it is not the case because of the existence of the
gravito-magnetic force.

In the FFF defined above, the total force exerted on the CoM of the BBH is
$m_1d^2\mathbf{x}_1/d\tau^2+m_2d^2\mathbf{x}_2/d\tau^2$. Now consulting
Equation~(\ref{eq:eom}), we can immediately eliminate the total force due to
Newtonian gravity because the gravitational forces on the two stellar-mass BHs
are of the same magnitude but in opposite directions.  As for the GEM force,
i.e., the second term on the right-hand side of Equation~(\ref{eq:eom}), we
first notice that the gravito-electric field, $\mathbf{E}_a$, is proportional
to the position $\mathbf{x}_a$ of an object. Therefore, the total
gravito-electric force, $\mathbf{F}_{\rm
CoM}^{E}=-m_1\mathbf{E}_1-m_2\mathbf{E}_2$, scales with
$m_1\mathbf{x}_1+m_2\mathbf{x}_2$, which is $m_{12}\mathbf{x}_{\rm CoM}$.
As a result,
if
at one instance we choose the origin of the FFF to coincide with the CoM of the
BBH, the total gravito-electric force will vanish. 
In summary, the self-gravity and the gravito-electric force will not induce
acceleration to the CoM if initially we place the BBH at $\mathbf{x}_{\rm
CoM}=0$ in the FFF. Consequently, these forces will not cause the outer orbit
to deviate from the geodesic of a test particle.

The effect of gravito-magnetic force is different because the force,
$-2m\mathbf{v}\times\mathbf{B}$, is not a linear function of the position
$\mathbf{x}$. For example, consider an initial
condition of $\mathbf{x}_{\rm CoM}=0$. This condition leads to 
$m_1\mathbf{B}_1+m_2\mathbf{B}_2=0$ because $\mathbf{B}_a$ is a linear function
of $\mathbf{x}_a$. From the above relationship, we further derive 
\begin{eqnarray}
	&&\mathbf{F}_{\rm CoM}^{B}=\sum_{a=1}^{2}
-2m_a(\mathbf{v}_a-{\boldsymbol\omega}\times\mathbf{x}_a)\times\mathbf{B}_a
	\\
	&&=2m_2
	{\boldsymbol\omega}\times(\mathbf{x}_2-\mathbf{x}_1)\times\mathbf{B}_2
-2m_2(\mathbf{v}_2-\mathbf{v}_1)\times\mathbf{B}_2,\label{eq:Bforce}
\end{eqnarray}
where $\mathbf{B}_2$ denotes the gravito-magnetic field at the position of
$\mathbf{x}_2$. The last equation shows that the total gravito-magnetic force
$\mathbf{F}_{\rm CoM}^{B}$ depends not on $\mathbf{x}_{\rm CoM}$, but on the
relative position $\mathbf{x}_2-\mathbf{x}_1$ and the relative velocity
$\mathbf{v}_2-\mathbf{v}_1$ of the two small BHs. Neither of them is zero.
Therefore, $\mathbf{F}_{\rm CoM}^{B}$ in general does not vanish.

Now we can understand why the CoM of the BBH does not follow a geodesic.  This
is because the non-vanishing gravito-magnetic force will inevitably drive the
CoM of the BBH away from the origin of the FFF,  even if initially
$\mathbf{x}_{\rm CoM}=0$.  Furthermore, a small displacement in
$\mathbf{x}_{\rm CoM}$ will be quickly amplified by the gravito-electric force
because the magnitude of the force scales with $m_{12}|\mathbf{x}_{\rm CoM}|$.
As the BBH revolves around the SMBH, the direction of the gravito-electric
force will, most of the time, point away from the origin of the FFF, causing
the CoM to displace further away from the geodesic of a test particle. 

\section{Numerical simulation}
\label{sec:num}

\subsection{Initial conditions}

Now we use numerical simulations to showcase the effects of the GEM forces on
the evolution of a BBH around a SMBH.  The parameters of the BHs are
$M=4\times10^6M_\odot$, $s=0.9$, $m_1=15M_\odot$, and $m_2=10M_\odot$.  

The semimajor axis of the BBH (inner bianry) is initially set to
$a_0=2,500m_{12}$, so that
the orbital period is $\tau_0=2\pi\sqrt{a_0^3/m_{12}}\simeq98~{\rm s}$ in its rest frame. Since
$a_0$ is much smaller than the curvature radius of the
background SMBH ($\sim r$), the precision in the calculation of
the GEM forces can be guaranteed. 
For simplicity, we assume that the inner orbit is initially circular. The
corresponding GW radiation timescale is $5a_0^4/(64m_1m_2m_{12})\simeq50$ yrs
\cite{peters_1964}, which is much longer than the mission duration of LISA.
The orbital plane is chosen to be coplanar with the outer orbit, which is
motivated by the prediction that BBHs could be delivered to the vicinities of
SMBHs by the accretion disks of AGNs \cite{peng21}.

For the outer orbit, we adopt the initial conditions of a test particle on a
circular, prograde orbit in the equatorial plane of the SMBH. Therefore, both
the inner and outer orbits are inside the equatorial plane. The orbital
semimajor axis is either $r=3.3M$ or $2.8M$, to cover the region of our interest while keeping the inner binary outside the ISCO and tidally stable. The corresponding 
orbital periods, measured by a distant observer, are $2\pi/\Omega\simeq866$ s
and $702$ s, where $\Omega$ is the orbital angular velocity in the
Boyer-Lindquist coordinates \cite{bardeen72}.

\subsection{Dynamics in the FFF}

The FFF, in which we will integrate Equation~(\ref{eq:eom}), is set up
according to the following two requirements. (i) The origin of the FFF
initially has the same position and velocity as the CoM of the BBH. (ii) The
$x$, $y$, and $z$ axes initially align with the $X^{r'}$, $X^{\theta'}$, and
$X^{\phi'}$ axes of the LIF (see Figure~\ref{fig:frames}).  Such a setup allows
us to use the method described in Section~\ref{sec:force} to simplify  the
calculation of the GEM forces.  Note that the initial conditions we chose
restrict the motion of the inner binary to the $x-z$ plane, although the method
is more general and can be used in misaligned cases. 

In our fiducial model, the outer orbit initially has a radius of $r=3.3M$. The angular
momentum of the inner binary points in the $-y$ direction, so that the inner
orbit rotates in the same sense as the rotation of the outer orbit. Following
the previous works \cite{yang19,secunda20,tagawa20,li21,li22}, we refer to this
orientation as the ``prograde'' one.  Viewed from the FFF, the inner binary
initially rotates much faster than the LIF. In fact, the angular velocity of
the inner binary,
$\omega_b=|\mathbf{v}_2-\mathbf{v}_1|/|\mathbf{x}_2-\mathbf{x}_1|$, is about
$6$ times greater than that of the LIF, $\omega$. Therefore, the second term in
Equation~(\ref{eq:Bforce}) predominates, and the CoM of the BBH will receive a
gravito-magnetic force. 

\begin{figure}[] 
\centering \includegraphics[width=0.3\textwidth]{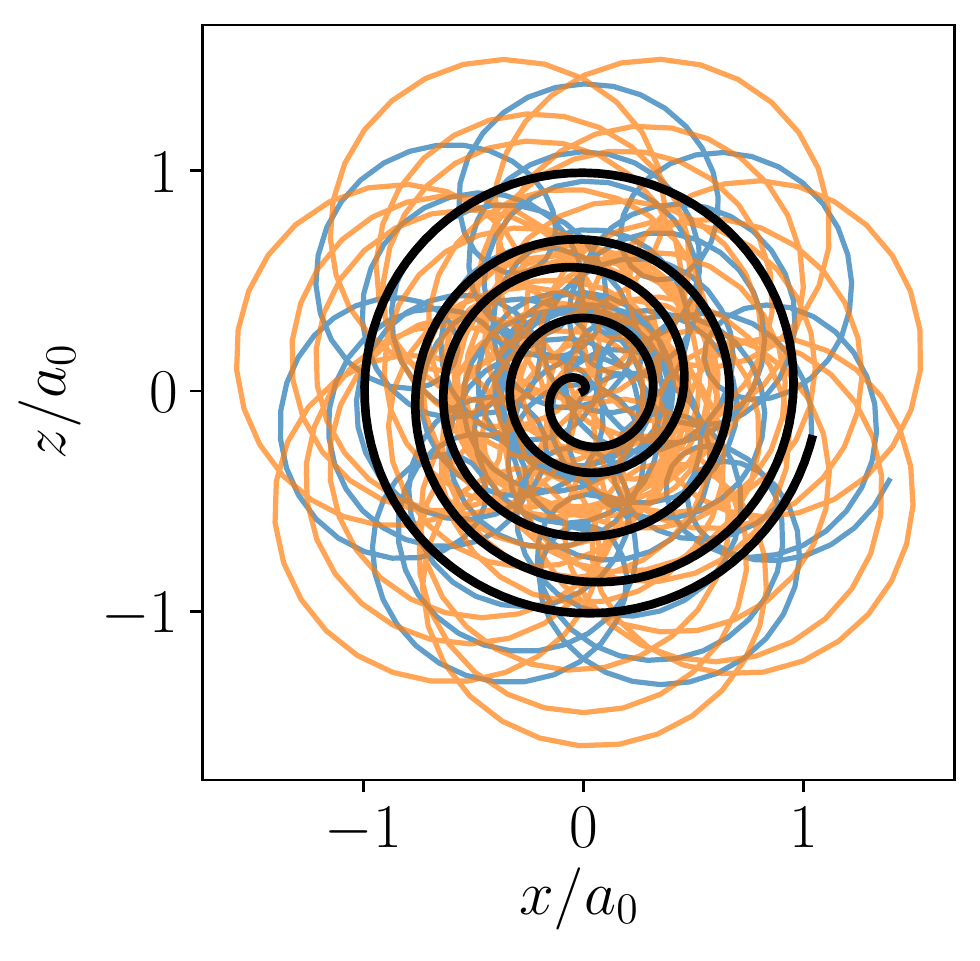}
	\caption{Evolution of a prograde BBH in the free-fall frame (FFF). 
The initial distance to the central SMBH is $r=3.3M$.
The blue and the orange
curves show the trajectories of the two BHs, and the black spiral curve corresponds to the center-of-mass of the BBH.}
\label{fig:R33fff}
\end{figure}

Figure~\ref{fig:R33fff} shows the evolution of the BBH in the FFF. The
trajectory of the CoM is shown as the black solid curve.  Initially, the CoM is
at the origin, $(x,z)=(0,0)$.  Then it is displaced by a small amount in the
positive-$x$ direction due to the non-vanishing gravito-magnetic force.
Afterwards, it spirals further away from the origin, which is driven mainly by
the gravito-electric force. The spiral pattern reflects the fact that the
$\mathbf{E}$ field is rotating in the FFF with a constant angular velocity of
$\omega$ due to the rotation of the LIF.  This result confirms the prediction
we made at the end of  Section~\ref{sec:deviation} that the outer orbit does
not follow a geodesic.

\subsection{Dynamics in the LIF}

\begin{figure}[] 
\centering \includegraphics[width=0.5\textwidth]{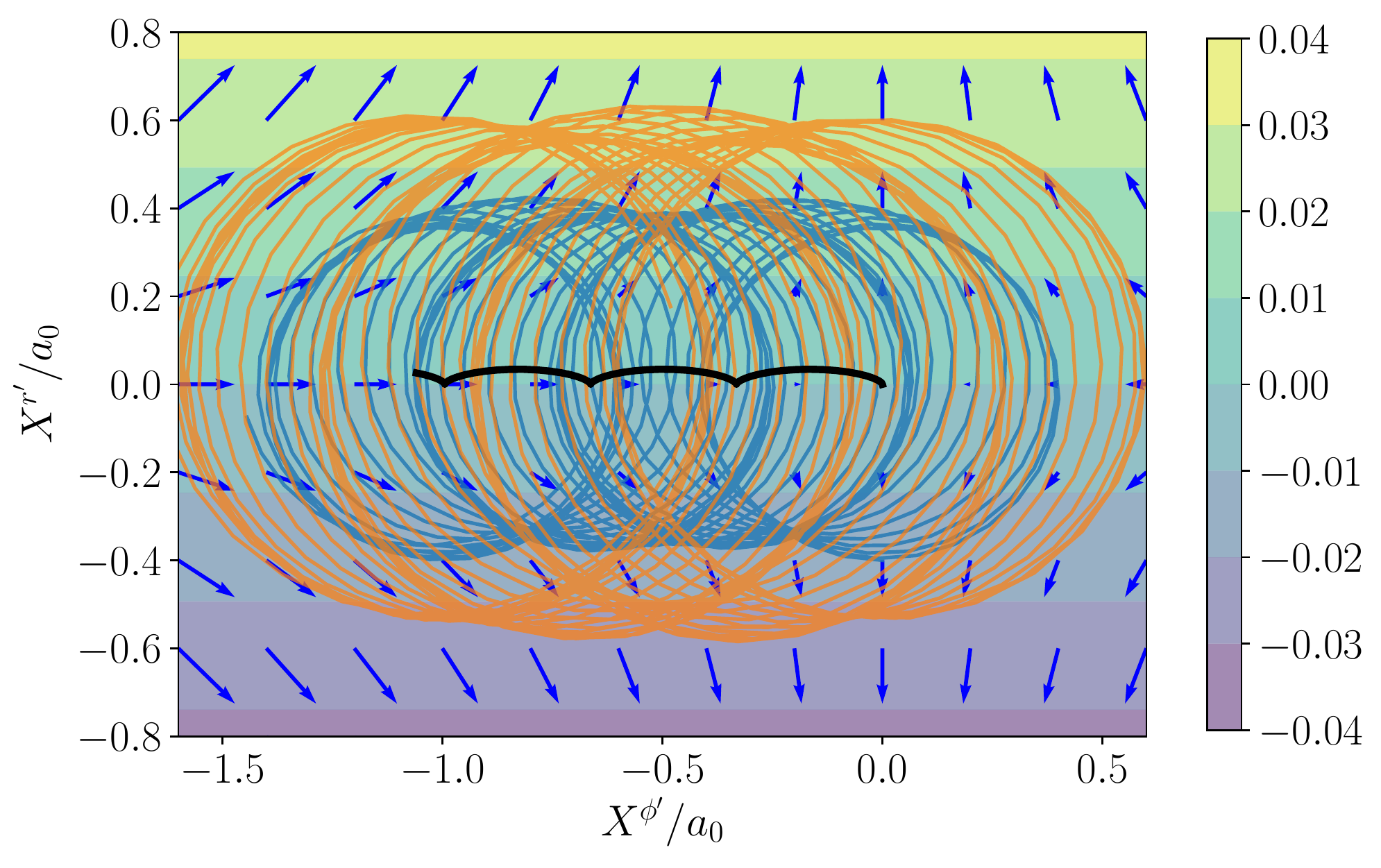}
\caption{Evolution of the prograde BBH in the local inertial frame (LIF). 
The blue, orange, and black curves have the same meanings as in Figure~\ref{fig:R33fff}. The blue
arrows show the direction of the gravito-electric force acting
on an object with unit mass . The length is proportional to the strength of the force. The background color indicates the strength
of the gravito-magnetic field, which is perpendicular to the $X^{r'}-X^{\phi'}$ plane according to the initial conditions of our model.}
\label{fig:R33lif}
\end{figure}

To see more clearly the directions of the GEM forces, we shown in
Figure~\ref{fig:R33lif} the evolution of the BBH in the LIF, where the GEM
fields do not rotate. In this frame, besides the self-gravity of the binary,
there are four types of additional forces acting on the BHs.

\begin{figure}[] 
\centering \includegraphics[width=0.48\textwidth]{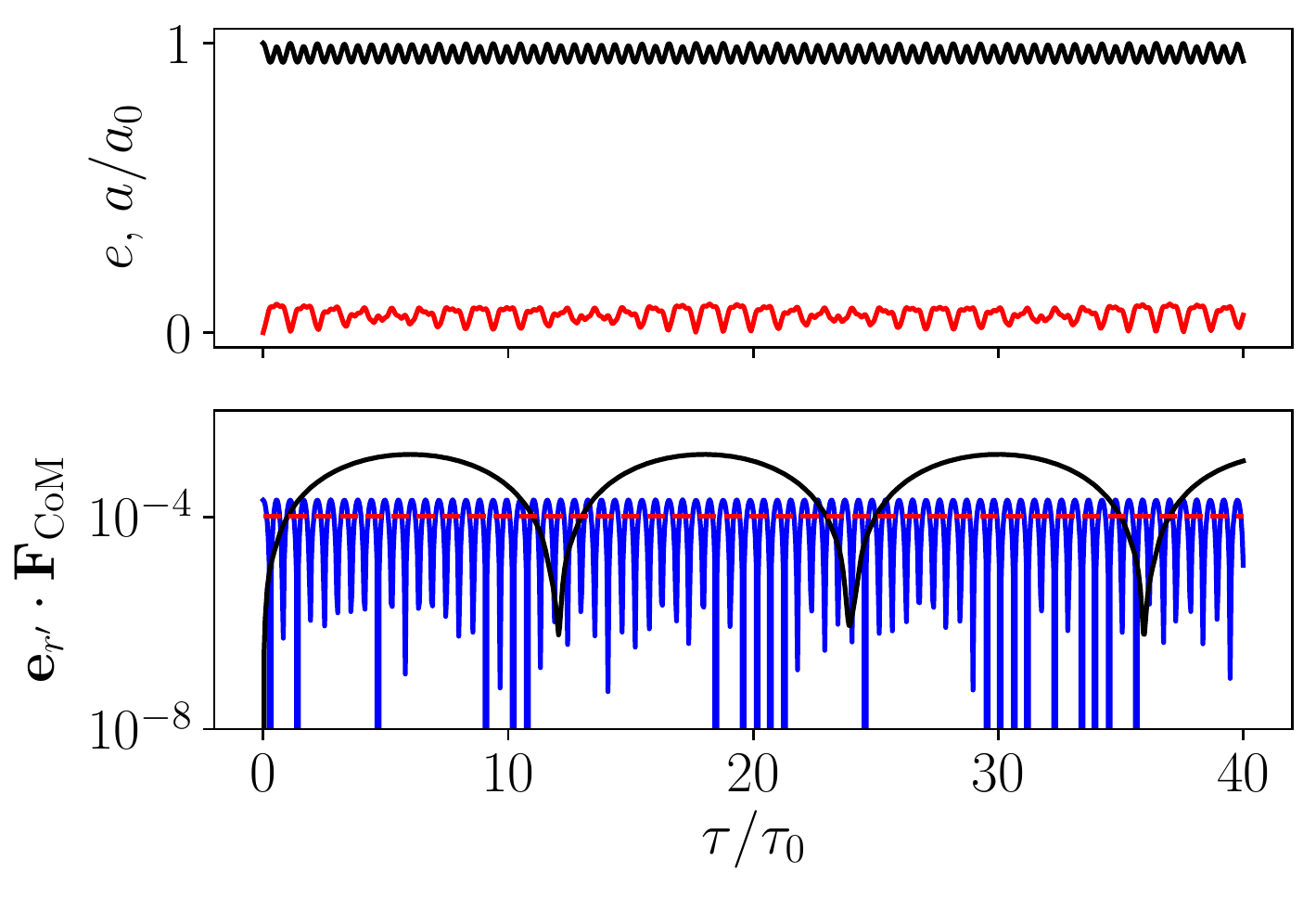}
\caption{Upper: Evolution of the semimajor axis (black line) and eccentricity (red line) of the
prograde BBH, viewed from the free-fall frame. Notice that time is in the unit of the initial orbital
period $\tau_0$ of the BBH.
Lower: Evolution of the radial components of the total gravito-magnetic 
	(blue curve)
and  gravito-electric (black curve) forces acting on the center-of-mass
of the BBH. The red dashed line is the time-averaged value of the above
gravito-magnetic force, computed according to Equation~(\ref{eq:BforceAve}). 
The values shown here are in simulation units.}
\label{fig:force}
\end{figure}

\begin{itemize}
\item[(i)] The gravito-electric force is shown as the blue arrows.  It
	vanishes only at the origin and increases with the distance from the
		origin. 

\item[(ii)] The gravito-magnetic force is induced by the gravito-magnetic
	field. The $\mathbf{B}$ field is parallel to the $X^{\theta'}$ axis
		(pointing out of the page) and its magnitude is indicated by
		the color. We can see that it is linearly proportional to
		$X^{r'}$, which is a direct result of the simplicity of the
		Riemann tensor in the LIF. 

To see the direction of the gravito-magnetic force, we 
notice that our BBH rotates clockwise in Figure~\ref{fig:R33lif}. Moreover, 
the binary remains relatively circular in our
fiducial model, as is suggested by the upper panel of Figure~\ref{fig:force}.
Furthermore, the sign of the $\mathbf{B}$ field 
changes when one moves from the upper half-plane in Figure~\ref{fig:R33lif}
into the lower one. Combining the above factors, we find that the gravito-magnetic force on either BH,
$-2m_a\mathbf{v}'_a\times\mathbf{B}_a$, always has a component pointing in the
positive-$X^{r'}$ direction, unless $X^{r'}=0$
in which case the $\mathbf{B}$ field vanishes.
Meanwhile, the $X^{\phi'}$ component of the gravito-magnetic force oscillates between
positive and negative values due to the rotation of the binary.  The oscillatory component averages out to zero
over one orbital period, but the $X^{r'}$ component does not. As a result,
the total gravito-magnetic force acting on the CoM of the BBH also has
a non-vanishing $X^{r'}$ component.

\item[(iii)]
A centrifugal force is also present because the LIF rotates with respect to the
FFF, the latter of which is an inertial frame globally. The magnitude of the centrifugal force, $m\omega^2\sqrt{(X^{r'})^2+(X^{\phi'})^2}$, is
also proportional to the distance to the origin, but compared with the
gravito-electric force, which is of the order of
$(m\gamma^2/r^2)\sqrt{(X^{r'})^2+(X^{\phi'})^2}$, the centrifugal force is
smaller by a factor of $\gamma^2(r/M)$, about $4.7$ in our fiducial model.

\item[(iv)] The rotation of the LIF also induces a Coriolis force,
$-m{\boldsymbol\omega}\times\mathbf{v}'$. It causes the velocity vector of a
moving object to rotate counterclockwise in Figure~\ref{fig:R33lif}.
\end{itemize}

The black solid curve in Figure~\ref{fig:R33lif} shows the trajectory of the
CoM of the BBH, which starts at the origin of the coordinates.
To understand its behavior, we first notice that the gravito-electric,
centrifugal, and Coriolis forces initially do not induce any acceleration on
the CoM of the BBH, because their values depend on either
$m_1\mathbf{x}_1+m_2\mathbf{x}_2$ or $m_1\mathbf{v}_1+m_2\mathbf{v}_2$, both
vanish according to our initial conditions.  

The only non-vanishing force acting on the CoM is the gravito-magnetic force.
We have mentioned above that it has a non-zero component
pointing in the positive-$X^{r'}$ direction.
To be more quantitative, we can calculate the $X^{r'}$ component of $\mathbf{F}^B_{\rm CoM}$
according to  Equation~(\ref{eq:Bforce}). Assuming that 
$|\mathbf{x}_{\rm CoM}|/a\simeq0$ and initially
the BBH is aligned with the $X^{r'}$ axis, we find that  
\begin{equation}
	\mathbf{e}_{r'}\cdot\mathbf{F}^B_{\rm CoM}=2\mu(\omega-\omega_b)a^2R_{t'r'r'\phi'}\cos^2\left[(\omega_b-\omega)\tau\right],
\end{equation}
where $\mathbf{e}_{r'}$ is the basis vector associated with the $X^{r'}$
axis,
$\mu=m_1m_2/m_{12}$ is the reduced mass of the binary, and we have
used the fact that in the LIF the $\mathbf{B}$ field only has a $X^{\theta'}$
component and the magnitude is $R_{t'r'r'\phi'}X^{r'}$.
Now we can see that this component is positive definite, unless the binary
is aligned with the $X^{\phi'}$ axis so that $\cos\left[(\omega_b-\omega)\tau)\right]=0$.
We show the value of this component in 
the lower panel of Figure~\ref{fig:force}
as the blue solid curve. The red dashed line in the same panel shows
the time-averaged value, 
\begin{equation}
	\left<\mathbf{e}_{r'}\cdot\mathbf{F}^B_{\rm CoM}\right>=\mu(\omega-\omega_b)a^2R_{t'r'r'\phi'}.\label{eq:BforceAve}
\end{equation}
Since it is positive, we can attribute the initial upward motion of the
CoM in Figure~\ref{fig:R33lif} to the $X^{r'}$ component of $\mathbf{F}^B_{\rm CoM}$. 

As soon as the CoM leaves the origin, the centrifugal force and, more importantly, the gravito-electric force (see the black solid curve in the lower panel
of Figure~\ref{fig:force})
start to take effect and further accelerate the CoM.  As the speed increases,
the Coriolis force, 
\begin{equation}
	\mathbf{F}_{\rm CoM}^{\rm C}=-2m_{12}{\boldsymbol\omega}\times\mathbf{v}'_{\rm CoM},
	\label{eq:Coriolis}
\end{equation}
becomes stronger and eventually exceeds the gravito-electric force. 
Since the Coriolis force makes the velocity vector to rotate, 
the
trajectory in Figure~\ref{fig:R33lif} bends over and finally comes back to the $X^{\phi'}$ axis.

Interestingly, as soon as the CoM is back on the $X^{\phi'}$ axis, its velocity
in the LIF vanishes.  This can be seen as a consequence of the conservation of
the Jacobian energy in a rotating frame.  More specifically, in our fiducial
model the inner binary remains circular so that neither its internal energy nor
the angular momentum significantly changes, as we have shown in the upper panel
of Figure~\ref{fig:force}. Therefore, the Jacobian energy of the outer orbit is
approximately a constant.  The vanishing of the velocities restores the initial
condition in the LIF.  As a result, the dynamical evolution described above
repeats. This is the reason that the black solid curve
in Figure~\ref{fig:R33lif} shows a periodic pattern.

\subsection{Evolution of the outer orbit}

\begin{figure}[] 
\centering \includegraphics[width=0.48\textwidth]{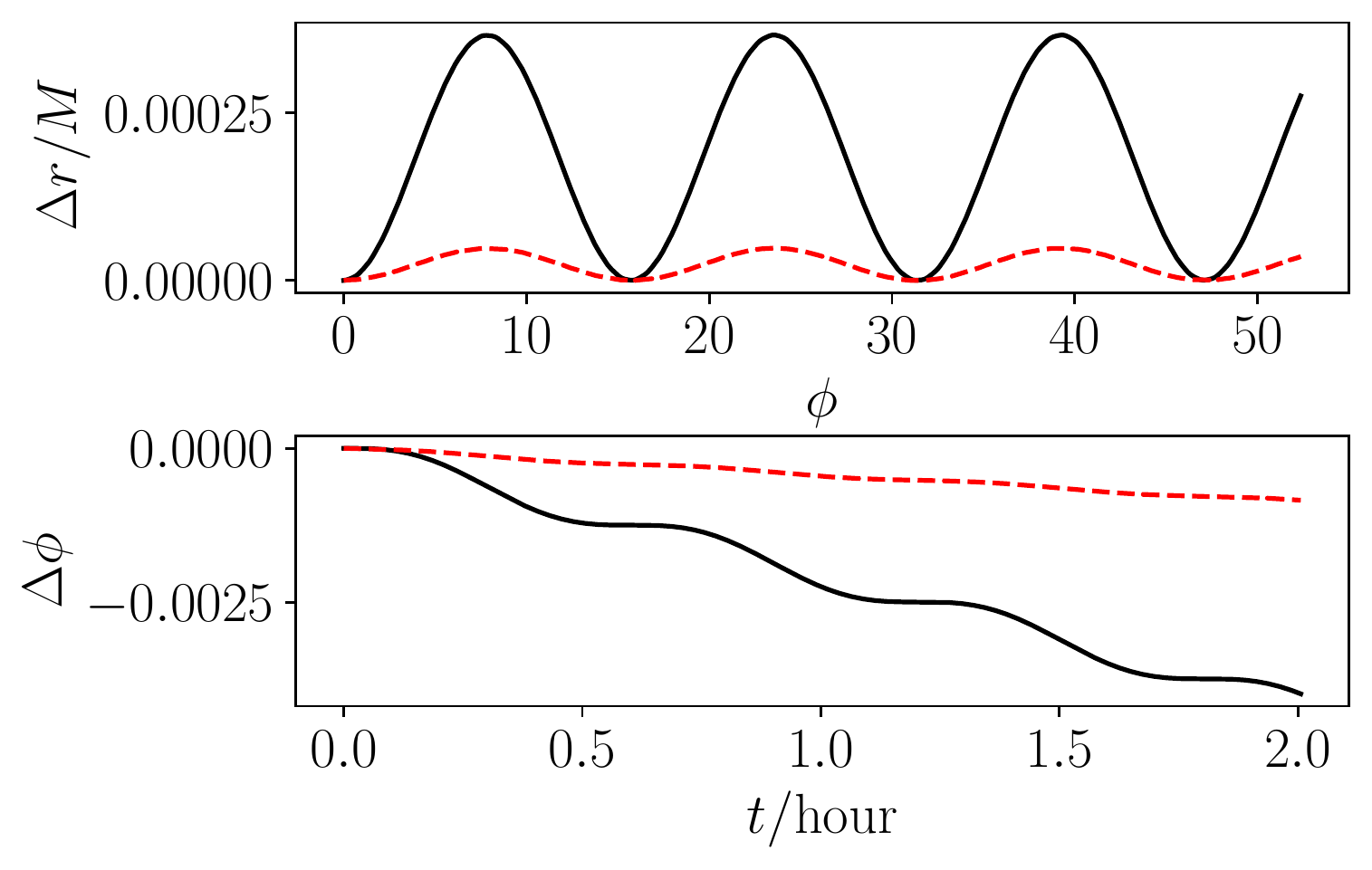}
\caption{Variation of the radial (upper panel) and azimuthal (lower panel)
	Boyer-Lindquist coordinates with respect to a circular geodesic. The 
	black curves refer to our fiducial model in which the CoM has no initial
	velocity in the FFF or LIF. The red dashed curves show the results when an initial
	velocity is added to the CoM to cancel the radial component
	of the gravito-magnetic force (see Section~\ref{sec:changeV}).}
\label{fig:OBL}
\end{figure}

The difference between the outer orbit and the geodesic of a test particle can
be more clearly seen in the Boyer-Lindquist coordinates, as is shown in
Figure~\ref{fig:OBL}.   The black solid curves correspond to the results from
our fiducial model.  From the upper panel, we find that the CoM periodically
takes excursions to slightly larger $r$ and comes back.  The period is much
longer than that of the outer orbit, as can be read from the values of the
azimuthal coordinate $\phi$ where one orbital period corresponds to $\Delta
\phi=2\pi$.  We can compare such a radial migration with the shrinkage of the
outer orbit due to GW radiation. In the former case, we have seen that the
radius has changed by a factor of $\Delta r/r\sim10^{-4}$  during one to two
revolutions of the outer orbit. In the latter case, where the GW radiation
timescale is approximately $(M/m_{12})(r/M)^{5/2}$ times the orbital period
\cite{peters_1964}, we find that the outer orbit in our fiducial model would
shrink by an amount of $\Delta r/r\sim10^{-6}$ during two orbital periods.
Therefore, the GEM force could dominate the radial evolution of the outer orbit
even when the system already enters the final stage of GW radiation.

The lower panel of Figure~\ref{fig:OBL} shows the difference in the orbital
phase. The black solid curve suggests that the CoM of our BBH is lagging behind
the test particle on a circular orbit. The phase difference is accumulative,
increasing by about $10^{-3}$ radian during each excursion of the CoM. As a
result, we would expect a phase difference of about $0.0025$ radian in one
hour, or as large as $1$ radian in about $17$ days. Such a deviation will
result in observable signatures, as we will discuss in Section~\ref{sec:obs}.

\section{Varying the initial conditions}\label{sec:initial}

\subsection{Velocity of the CoM}\label{sec:changeV}

We have seen that if the outer orbit starts on a circular geodesic, it will
soon deviate from circular motion by oscillating radially. The oscillation is
driven initially by the gravito-magnetic force, which on average accelerates
the CoM of the BBH along the radial direction. 

We notice that such a force can be balanced by the Coriolis force if we relax
the initial condition, giving the CoM of the BBH an extra velocity in the
$X^{\phi'}$ (azimuthal) direction. Introducing such a Coriolis force, in
principle, could reduce the radial acceleration and hence suppress the
amplitude of the radial oscillation. In this way, a circular orbit may be
sustained. 

To test the above postulation, we give the CoM an initial velocity along the
$X^{\phi'}$ axis ($z$ axis in the FFF).  The magnitude of this velocity is
given by equating Equations~(\ref{eq:BforceAve}) and (\ref{eq:Coriolis}). To
transfer this velocity into the FFF, we use the fact that initially
$\mathbf{v}'_{\rm CoM}=\mathbf{v}_{\rm CoM}$ because $\mathbf{x}_{\rm CoM}=0$.

The results are shown in Figure~\ref{fig:OBL} as the red dashed curves.  The
upper panels shows that, indeed, the radial oscillation is significantly
smaller than that in the previous case, where the CoM has no initial velocity
in the azimuthal direction. However, the amplitude of the oscillation is not
zero. This can be explained by the fact that the gravito-magnetic force is
oscillatory but the Coriolis force is varying more smoothly.  Therefore, an
initial azimuthal velocity could reduce, but not completely suppress the radial
oscillation of a binary around a SMBH. 

The red dashed curve in the lower panel of Figure~\ref{fig:OBL} shows that the
phase shift is also much smaller when an azimuthal velocity is initially added
to the CoM of the BBH. Therefore, detecting such BBH would require much longer
observing time.

\subsection{Orientation of the inner orbit}

In our fiducial model, the inner binary initially rotates in the same direction
as the outer orbit. We have seen that in such a prograde case, the CoM of the
inner binary will drift in the negative azimuthal direction. If we flip the
rotation axis of the inner binary to make it counter-rotating with respect to
the outer orbit, according to Equation~(\ref{eq:Bforce}), the dominant part of
the gravito-magnetic force, i.e., the term proportional to
$\mathbf{v}_2-\mathbf{v}_1$, will change sign. Then the CoM of the inner binary
initially accelerates in the opposite direction relative to the fiducial case.
We expect that the subsequent azimuthal drift of the CoM will also
flip direction.

\begin{figure}[] 
\centering \includegraphics[width=0.5\textwidth]{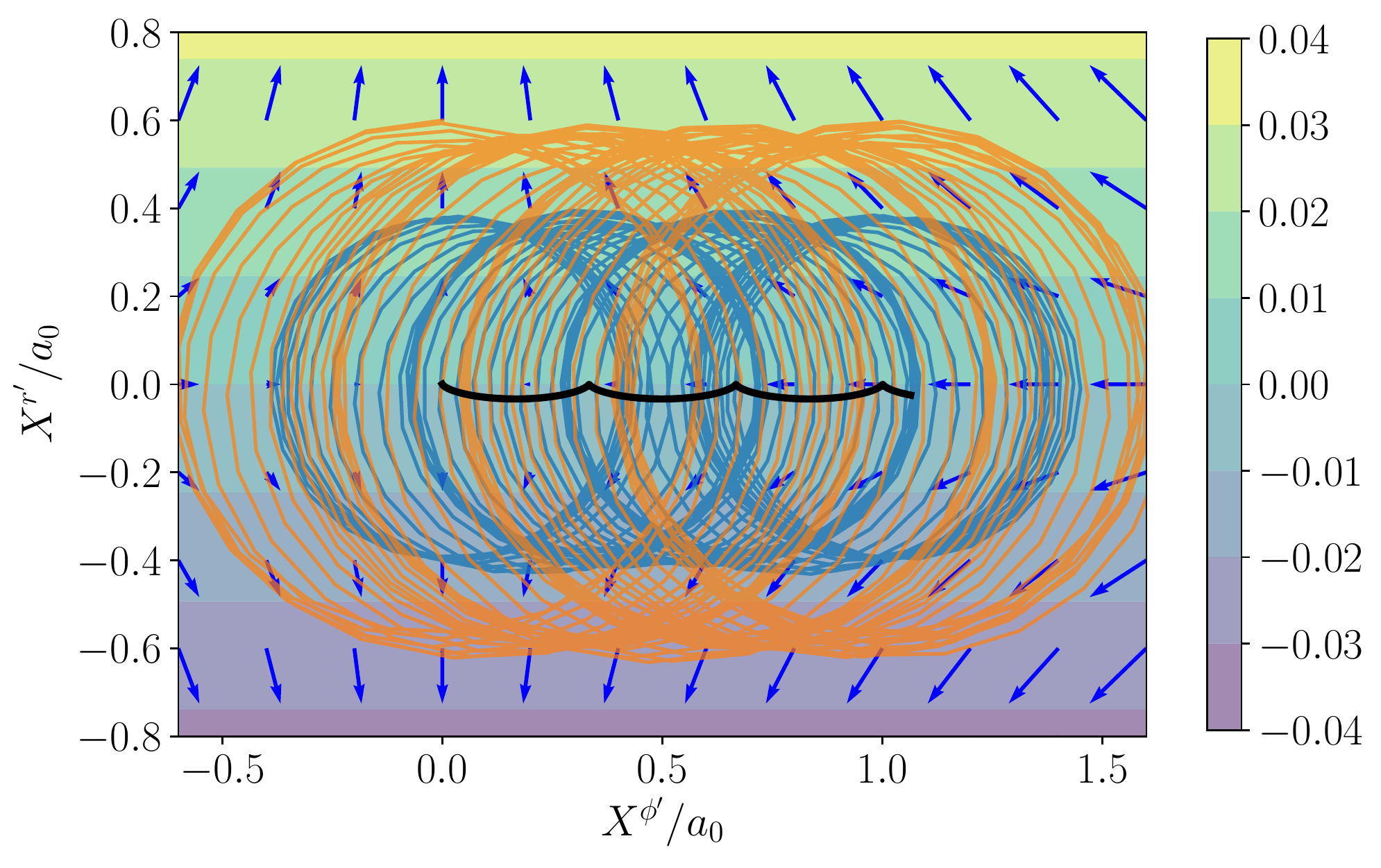}
\caption{Same as Figure~\ref{fig:R33lif} but the BBH is rotating in the 
	retrograde direction, i.e., the angular momentum is pointing
	out of the page. The CoM of the BBH starts from the origin,
	$(X^{r'},X^{\phi'})=(0,0)$.}
\label{fig:R33lifretro}
\end{figure}

\begin{figure}[] 
\centering \includegraphics[width=0.48\textwidth]{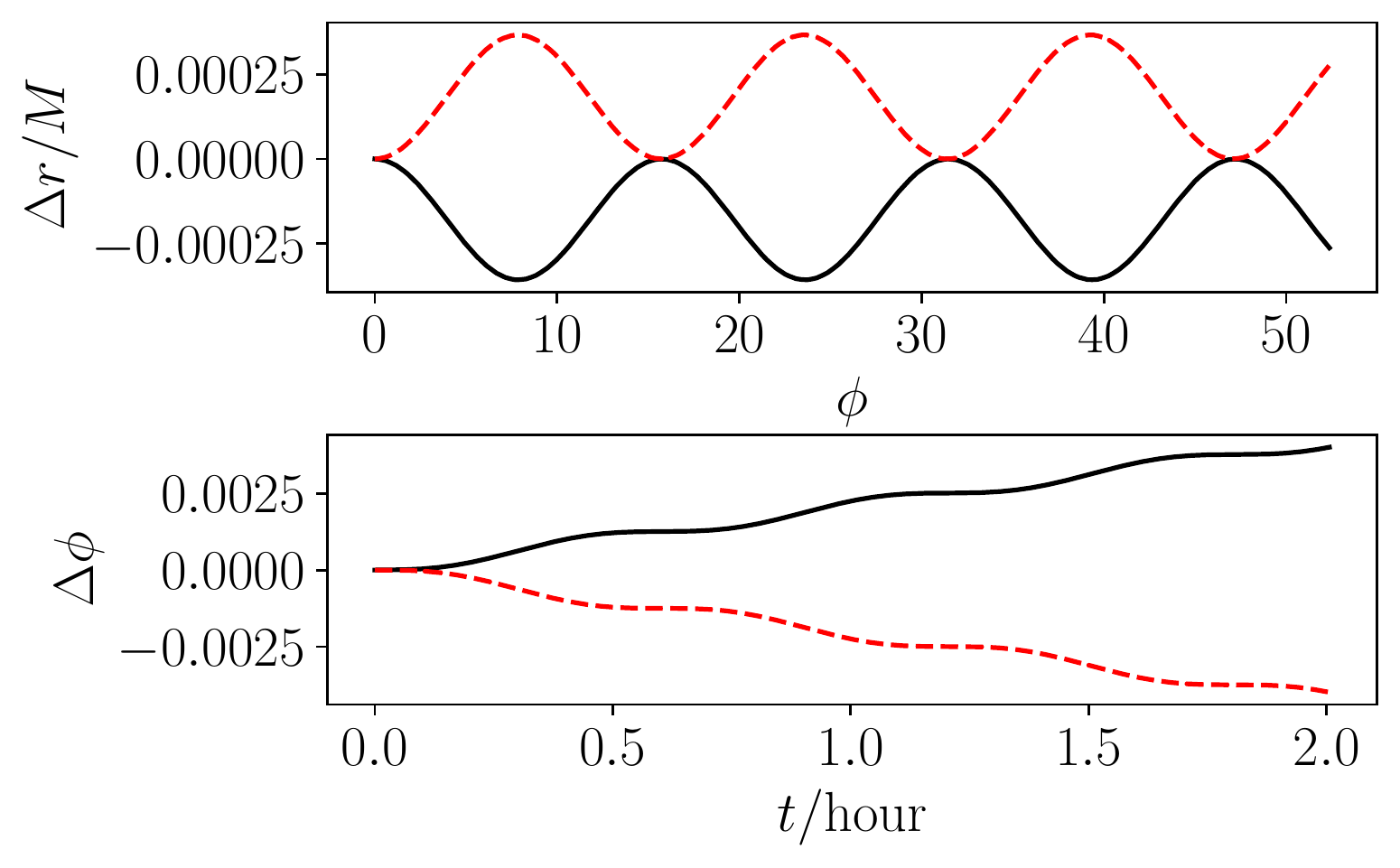}
\caption{Same as Figure~\ref{fig:OBL} but only showing the simulations without
an initial CoM velocity in the FFF or LIF. The black solid and red dashed
curves, respectively, refer to the simulations with
retrograde and prograde BBHs.
}
\label{fig:OBLretro}
\end{figure}

Figure~\ref{fig:R33lifretro} shows the result from our simulation of a
retrograde inner binary. Comparing with the trajectory of the CoM shown in
Figure~\ref{fig:R33lif}, we find that the inner binary indeed reverses the
drift direction when its rotation axis flips.  Figure~\ref{fig:OBLretro}
compares the Boyer-Lindquist coordinates of the CoM in the prograde
and retrograde cases. We can see that in both the radial (upper panel) and the
azimuthal directions (lower panel), the direction of the drift flips as the
sign of the angular momentum of the inner binary changes.

\subsection{Distance to the SMBH}

To study the evolution of the BBH at a smaller distance to the SMBH, we reduce
the initial radius of the outer orbit to $r=2.8M$ and rerun the simulation. The
other parameters are kept the same as those in the fiducial model.

\begin{figure*}
\centering
\begin{minipage}{.3\textwidth}
  \centering
  \includegraphics[width=\linewidth]{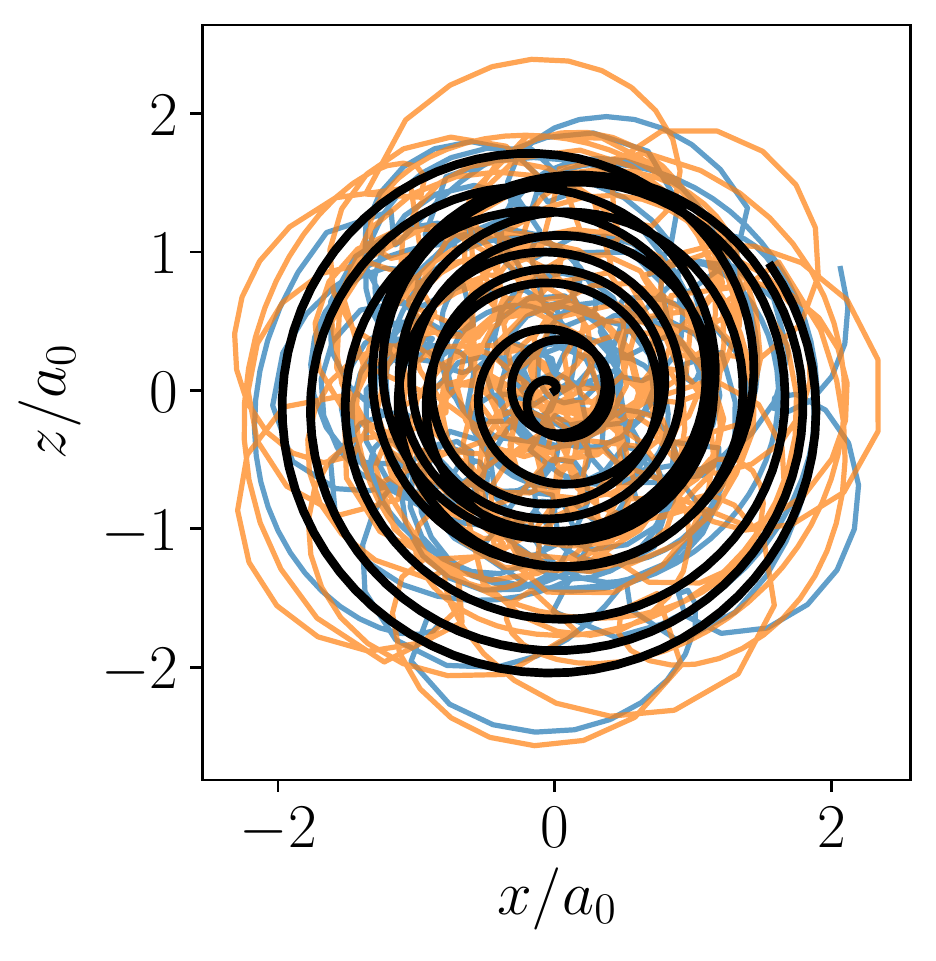}
\end{minipage}%
\begin{minipage}{.5\textwidth}
  \centering
  \includegraphics[width=\linewidth]{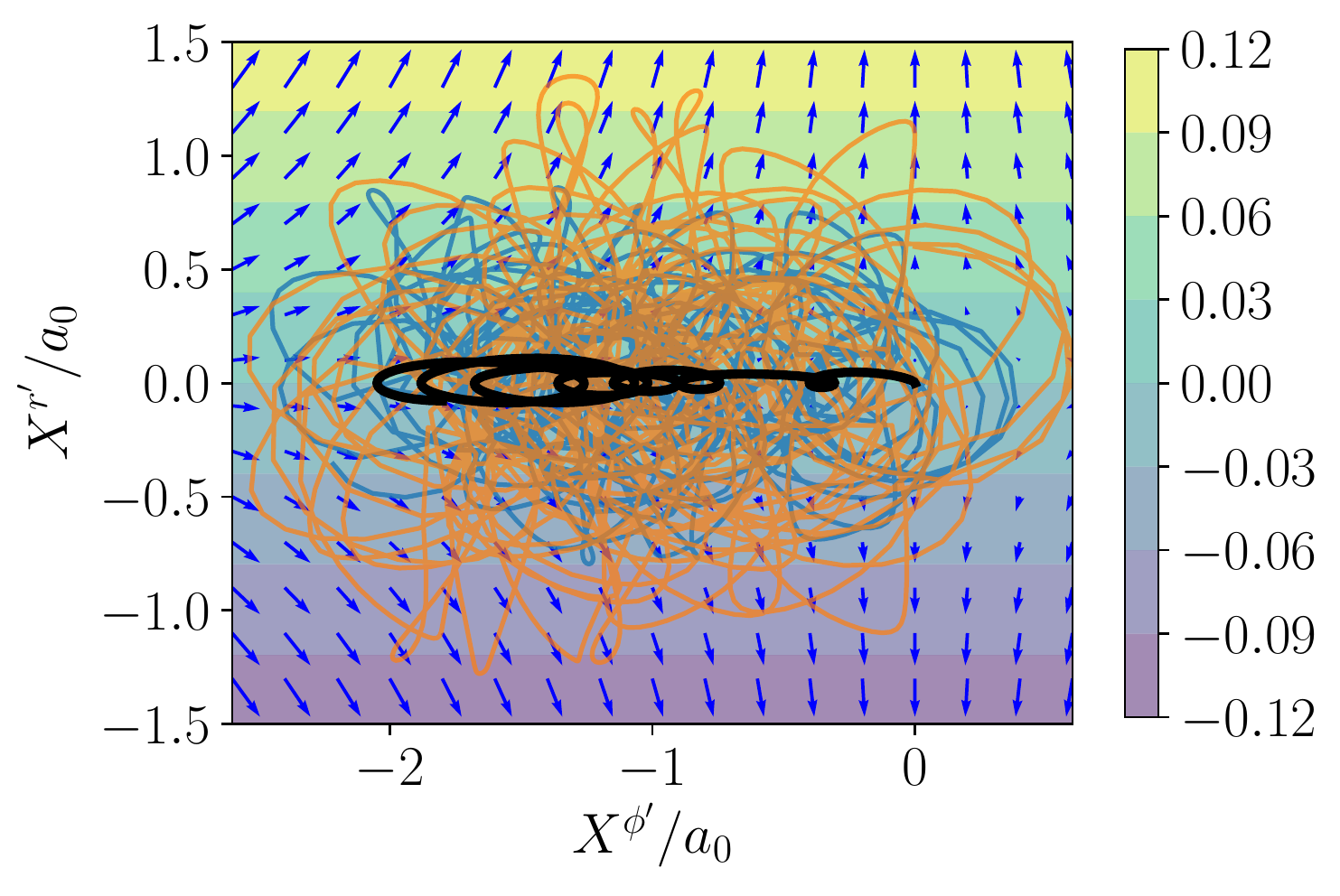}
\end{minipage}
\caption{Evolution of a prograde BBH when the initial distance
to the SMBH is $r=2.8M$. The left panels shows the evolution
in the FFF and the right one in the LIF. The lines, arrows, and colors have the same
meanings as in Figures~\ref{fig:R33fff} and \ref{fig:R33lif}.
}
  \label{fig:R28}
\end{figure*}

Figure~\ref{fig:R28} shows the resulting trajectories of the two stellar mass
BHs and their CoM in the FFF (left) and the LIF (right panel).  We can see that
the inner binary is no longer circular.  The trajectory of the CoM is also more
irregular compared with that in the fiducial model. In particular, the
direction of its azimuthal drift reverses several times, as can be seen in the
LIF. These behaviors are qualitatively different compared with that at a larger
distance from the SMBH.

\begin{figure}[] 
\centering \includegraphics[width=0.48\textwidth]{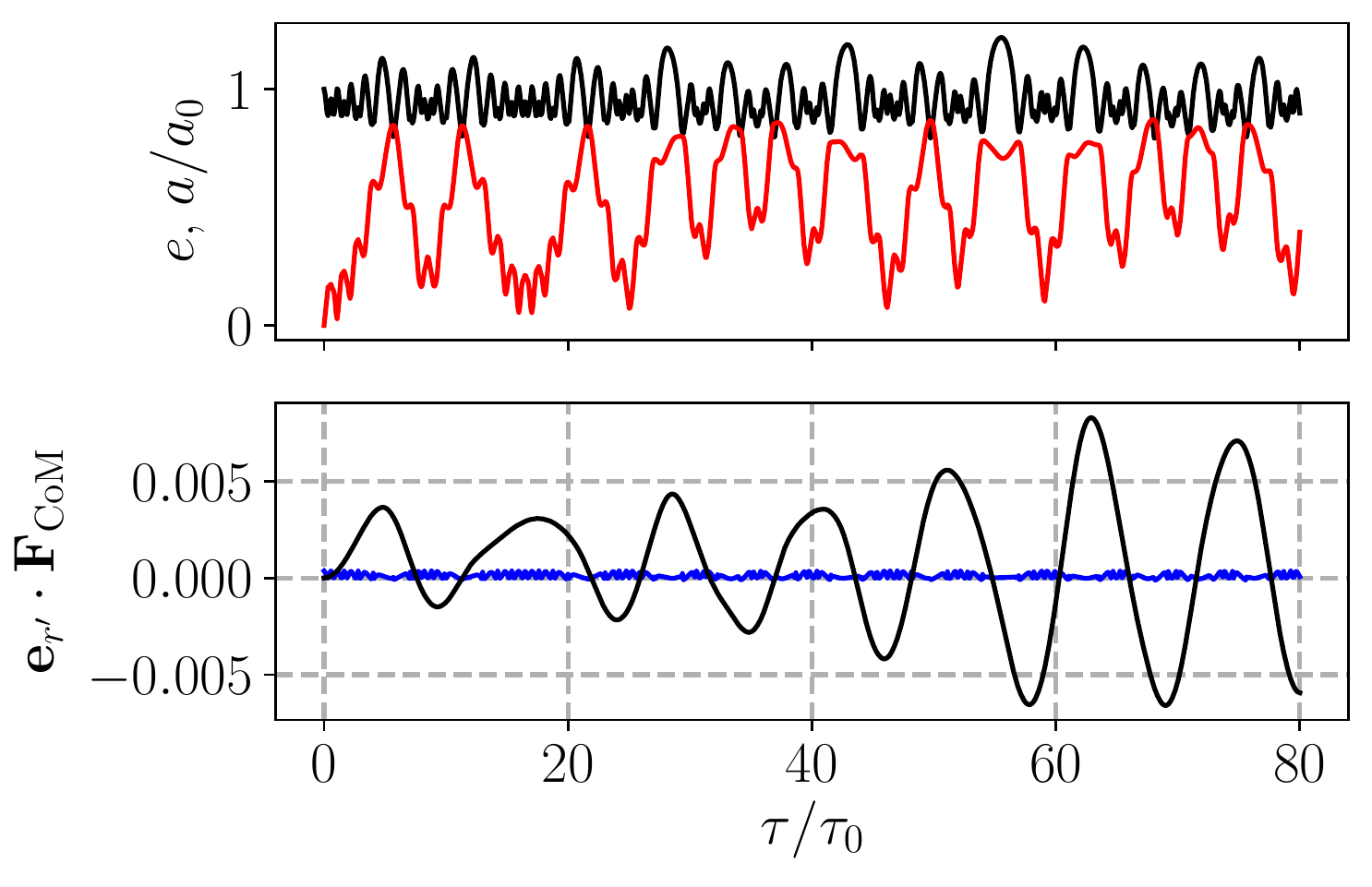}
\caption{The same as Figure~\ref{fig:force} but for $r=2.8M$.}
\label{fig:force28}
\end{figure}

To understand the cause of the irregularity, we show in the upper panel of
Figure~\ref{fig:force28} the evolution of the semimajor axis and the
eccentricity of the inner binary.  Now both parameters vary with significantly
larger amplitude compared with the variation in the fiducial model
(Figure~\ref{fig:force}).  The larger amplitude is a result of the stronger GEM
forces at smaller $r$.  Furthermore, we find that the eccentricity of the inner
binary is quasi-periodically excited to a value as large as $0.9$, on a
timescale roughly ten times longer than the initial orbital period of the inner
binary ($\tau_0$). The evolution 
does not resemble the Von Zeipel-Lidov-Kozai
cycle not only because of its irregularity, but also because
the Von Zeipel-Lidov-Kozai
cycle requires 
the inner and outer orbits to be misaligned, but they are coplanar in our case.

The irregularity of the inner orbit also makes the GEM forces on individual BHs
more complex. As is shown in the lower panel of Figure~\ref{fig:force28},
neither the gravito-magnetic nor the gravito-electric force behaves according
to a simple sinusoidal curve. However, the $X^{r'}$ component of the
gravito-magnetic force remains positive, reflecting the fact that the inner
binary remains rotating in the prograde direction. On the contrary, the
$X^{r'}$ component of the gravito-electric force could reach large negative
values during the evolution, suggesting that the CoM of the BBH has wandered
far into the lower half-plane of the LIF.

\begin{figure}[] 
\centering \includegraphics[width=0.48\textwidth]{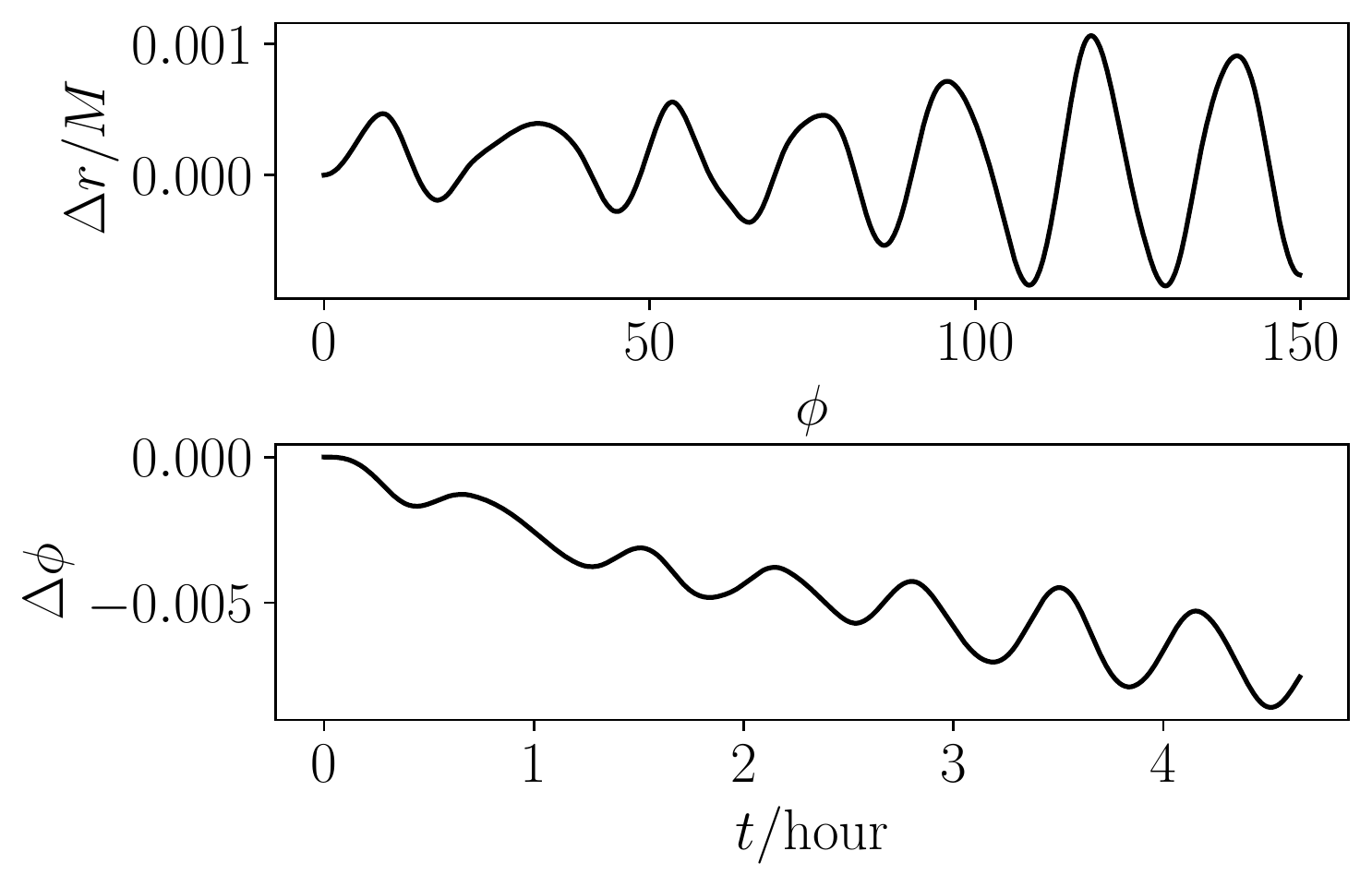}
\caption{The same as Figure~\ref{fig:OBL} but for $r=2.8M$.}
\label{fig:OBL28}
\end{figure}

Figure~\ref{fig:OBL28} shows the variation of the Boyer-Lindquist coordinates
of the outer orbit. Comparing it with the results from the fiducial model, we
find that the variation in the radial direction has larger amplitude now. More
importantly, the amplitude seems to increase with time. This behavior can be
understood as the consequence of an exchange of energy between the inner BBH
and the outer orbit. Moreover, the phase shift in the azimuthal direction,
$\Delta \phi$, also accumulates much more quickly when $r$ is smaller.  More
quantitatively, the magnitude of $\Delta \phi$ could reach $0.005$ radian in
about two hours, which amounts to $1$ radian in about eight days.

Because of its irregular variation, the gravito-magnetic force can no longer be
counter-balanced by a constant Coriolis force in the LIF.  Therefore, we find
that introducing an initial velocity to the CoM will not reduce the amplitude
of the radial or azimuthal drift in the LIF.  For this reason, we do not show
the corresponding results in Figure~\ref{fig:OBL28}.

\section{Observable signatures}\label{sec:obs}

In our model, we have chosen an inner binary with an orbital frequency of
about $10$ milli-Hertz (mHz) in its rest frame. Taking into account the facts
that the GW frequency of a circular binary is twice the orbital frequency and
the observed frequency would be further modulated by a factor of a few due to
the gravitational and Doppler redshifts \cite{chen_li_2019}, we find that the
inner binary, i.e., the BBH, remains in the sensitive band of LISA.

The GW signal from such an inner binary has been studied in detail in recent
years. These earlier studies have revealed a significant shift of the GW phase
which can be used to distinguish the BBHs around SMBHs (see a summary in
Section~\ref{sec:intro}). Here we point out an additional phase shift
based on the results shown Figures~\ref{fig:force} and \ref{fig:force28}. We
can see that the semimajor axis and eccentricity of the inner binary oscillate
rapidly.  The oscillation of these orbital elements will affect how the orbital
phase, and hence the GW phase, increases with time.  The associated timescale
is the rotation period of the BBH, which is much shorter than the period of the
outer orbit or the Von Zeipel-Lidov-Kozai cycle.  The short timescale is
closely associated with the driving mechanism, which is the variation of the
GEM forces during the mutual rotation of the two small BHs.  
Therefore, detecting a phase shift on a timescale comparable to the
period of the GWs might indicate that the BBH is close to a SMBH.

The outer orbit of our system, seen by a distant observer, has a frequency of
about $1$ mHz. Therefore, the GWs emitted by the outer orbit may also be
detectable by LISA.  While the previous studies of similar triple systems have
modeled the outer orbit with a geodetic motion (free fall), our work clearly
shows that the orbit is not in free fall. In principle,  the deviation is
detectable by contrasting the observed signal with the template of a standard
EMRI, i.e., an EMRI with only one stellar-mass BH whose mass is $m_{12}$. A
mismatch by one radian would be sufficient to reveal the peculiarity of the
small object around the SMBH. Finding such a mismatch needs only one to two
weeks of data, according to the phase shifts shown in Figures~\ref{fig:OBL},
\ref{fig:OBLretro}, and \ref{fig:OBL28}.  In practice, distinguishing such a
phase shift from the effect caused by other environmental factors, such as a
perturbation by the surrounding gas or stars
\cite{Amaro-SeoaneBremCuadraArmitage2012,barausse14,derdzinski21,zwick21,zwick22},
might be difficult and requires further investigation. 

Another interesting signature revealed by our simulations is that the direction
of the phase shift of the outer orbit is correlated with the orientation of the
inner binary. As is shown in the lower panel of Figure~\ref{fig:OBLretro}, a
prograde inner binary drifts along the negative azimuthal direction, while a
retrograde one drifts towards the positive azimuthal direction. Such a
difference can be used in future EMRI observations to infer the orientations of
the inner binaries. According to the previous studies, knowing the orientation
would help us understand the dynamical processes responsible for the production
of stellar-mass BHs around SMBHs \cite{yang19,secunda20,tagawa20,li21,li22}.

\section{Conclusions and future work}\label{sec:con}

In this study, we have investigated the evolution of a BBH at a distance of
only $2-3$ gravitational radii from a Kerr SMBH.  We showed that such a highly
relativistic triple can be modeled using Newtonian dynamics plus a perturbative
GEM force if one takes advantage of its hierarchy and investigate the dynamics
in a frame freely falling alongside the BBH.  Our main finding is that the CoM
of the BBH does not follow a geodesic line.  We identified the cause of the
geodetic deviation to be a non-vanishing GEM force on the CoM. We pointed out
that within several weeks, the deviation will be large enough to be detectable
by LISA.  Find such a signal could reveal the binarity of the small object in
an EMRI source, as well as put constraint on the orientation of the inner
binary.

So far, we have restricted ourselves to the systems with near-circular outer
orbits. Although this choice is motivated by a class of astrophysical models,
our method in principle can be applied to elliptical, parabolic, or hyperbolic
outer orbits with a slight modification, e.g., by choosing an appropriate
free-fall observer on the same orbit. Such an improvement is useful because
non-circular orbits could also be populated with BBHs due to a dynamical
process called ``tidal capture''
\cite{2018CmPhy...1...53C,addison_gracia-linares_2019}.

Our current method is insufficient to track the BBH when it wanders
substantially far away from the origin of the FFF. This is because when the
separation between the origin of the coordinates and the CoM of the BBH is
comparable to $M$, the perturbation induced by the background Kerr SMBH to the
metric of the FFF will no longer be small. However, as long as the size of the
BBH remains small relative to $M$, we can circumvent the problem  by
resetting the FFF, making it centered on the BBH. The accuracy of such a method needs to be tested. 

We have not included the effect of GW radiation in our simulations.  For the
inner BBH, we can include the effect by adding PN corrections to the equation
of motion since the semimajor axis $a$ is much larger than $m_{12}$. These
corrections will become important when the BBH obtains high eccentricity, as we
have seen in the case of a small distance between the BBH and the SMBH
($r=2.8M$). In particular, the dissipative PN terms will drive the BBH to
coalesce when $e$ is large. Therefore, we will be able to test the correlation
between a small $r$ and a higher probability of BBH coalescence, as was
envisioned in \cite{chen_li_2019}.  For the outer orbit, the GW radiation will
lead to a gradual shrinkage of $r$.  It might result in an interesting
situation in which the effect of GW radiation is counter-balanced by the
increase of $r$ as we have seen in Figures~\ref{fig:OBL} and \ref{fig:OBL28}.
This could be a new way of trapping BBHs at the last few gravitational radii
from a SMBH, in addition to the mechanism proposed in a previous work
\cite{peng21}.

Although there is still much space for improvement, the current work has
established a practical framework to fast generate the orbit of a BBH very
close to a SMBH ($r\lesssim10M$) while keeping most of the essential
relativistic effects included. This framework will also enable us to build more
accurate waveform templates, which are crucial to the detection of such triple
systems in future GW missions like LISA. 

\section{Acknowledgement}

This work is supported by the National Science Foundation of China grants No.
11991053 and 11873022, and the National Key Research and Development Program of
China Grant No. 2021YFC2203002.  The authors would like to thank Zhoujian Cao
and Bing Sun for their help in the problem of  coordinate transformation.

\bibliographystyle{apsrev4-1.bst}
\bibliography{aamnem99,biblio,bibbase}

\end{document}